\documentclass[12pt]{article}
\usepackage{amsfonts}
\usepackage{eurosym}
\usepackage{graphicx}
\usepackage[utf8]{inputenc}
\usepackage[T1]{fontenc}
\usepackage{indentfirst}
\usepackage[top=0.6in,bottom=0.6in,left=0.6in,right=0.6in]{geometry}
\usepackage[final]{hyperref}
\usepackage{amsmath}
\usepackage{hyperref}
\usepackage{cite}
\usepackage{subcaption}
\usepackage{caption}
\usepackage{amssymb}
\usepackage{multirow}
\usepackage[table]{xcolor}
\usepackage{orcidlink}
\usepackage{authblk}
\usepackage{float}

\hypersetup{
colorlinks=true,
linkcolor=blue,
citecolor=blue,
filecolor=magenta,
urlcolor=blue
}

\begin{document}

\title{Generalized Dunkl Quantum Systems with Energy-Dependent Interactions: Exact Solvability and Thermodynamic Properties}
\author[1]{B. Hamil \orcidlink{0000-0002-7043-6104} \thanks{hamilbilel@gmail.com, bilel.hamil@umc.edu.dz (Corresponding author)}}
\author[1]{R. Boudjouraf \orcidlink{0009-0005-7200-059X} \thanks{radhia.boudjouraf@doc.umc.edu.dz, radiaboudj56@gmail.com}}

\author[1]{A. Benchikha \orcidlink{0000-0001-6467-5005} \thanks{benchikha4@yahoo.fr, amar.benchikha@umc.edu.dz}}
\author[2]{M. Merad \orcidlink{0000-0001-7547-6933} \thanks{meradm@gmail.com}}

\affil[1]{Laboratoire de Physique Mathématique et Subatomique (LPMPS), Faculté des Sciences Exactes, Université Constantine 1-Frères Mentouri, Constantine, Algeria.}
\affil[2]{Laboratoire de systèmes dynamiques et contrôle (L.S.D.C), Département des sciences de la matière, Faculté des Sciences Exactes et SNV, Université de Oum-El-Bouaghi, 04000, Oum El Bouaghi, Algeria.}
\date{}
\maketitle

\begin{abstract}
This work investigates a class of exactly solvable quantum systems described by a generalized Dunkl–Schr\"{o}dinger equation with energy-dependent interactions. A two-parameter extension of the Dunkl differential operator is introduced, incorporating both the conventional reflection contribution and an additional derivative-reflection coupling. The resulting formalism leads to a modified Heisenberg algebra, from which the continuity equation, conserved probability current, and generalized normalization condition are derived for energy-dependent potentials. As an application, an energy-dependent harmonic oscillator is studied in detail. Exact analytical expressions for the eigenfunctions and energy spectrum are obtained in terms of associated Laguerre polynomials. The combined effects of the Wigner parameter, the deformation parameter, and the energy-dependence parameter are shown to produce parity splitting, nonlinear energy spectra, and nonuniform level spacing while preserving exact solvability. The thermodynamic properties of the model are then examined through the canonical partition function using the Euler–Maclaurin summation formula. Analytical expressions for the partition function, internal energy, entropy, and specific heat are derived and analyzed numerically. The results demonstrate that the generalized Dunkl deformation and energy-dependent interaction provide effective control over both the spectral and thermal characteristics of the system, extending the class of exactly solvable quantum models with reflection symmetry.
\end{abstract}

\affil[1]{Laboratoire de Physique Mathématique et Subatomique (LPMPS),
Faculté des Sciences Exactes, Université Constantine 1-Frères Mentouri,
Constantine, Algeria.} 
\affil[2]{Laboratoire de systèmes dynamiques et
contrôle (L.S.D.C), Département des sciences de la matière, Faculté des
Sciences Exactes et SNV, Université de Oum-El-Bouaghi, 04000, Oum El
Bouaghi, Algeria.}

\affil[1]{Laboratoire de Physique Mathématique et Subatomique, Faculté des
Sciences Exactes, Université Constantine 1 Frères Mentouri, Constantine,
Algeria.} 
\affil[2]{Laboratoire de systèmes dynamiques et contrôle
(L.S.D.C), Département des sciences de la matière, Faculté des Sciences
Exactes et SNV, Université de Oum-El-Bouaghi, 04000, Oum El Bouaghi,
Algeria.}

\section{Introduction}

\label{sec1} Quantum mechanics is formulated through the canonical quantization
procedure, where classical observables are converted into operators acting
on a Hilbert space, and canonical Poisson brackets are replaced by
commutation relations. In particular, the position and momentum operators
are required to satisfy: 
\begin{equation*}
\left[ x,p\right] =i\hbar .
\end{equation*}

This relation serves as the basis for the Heisenberg representation, where
the time evolution of observables is governed entirely by the Heisenberg
equations of motion. While this framework has been exceptionally successful
in describing microscopic phenomena, modern quantum field theory still
relies on mathematical workarounds, such as regularization and
renormalization whose conceptual foundations remain a subject of debate.
These persistent difficulties suggest that the current formal structure may
be an effective description rather than the final form of the theory,
motivating the search for alternative formulations that remain compatible
with established physical principles.

This raises a fundamental question: can the canonical commutation relations
be derived from the equations of motion? Eugene P. Wigner first addressed
this question in his seminal 1950 paper \cite{Wigner}, \textsl{Do the
Equations of Motion Determine the Quantum Mechanical Commutation Relations?}
By analyzing the one-dimensional harmonic oscillator, Wigner demonstrated
that the equations of motion admit an infinite family of compatible
commutation relations, with the canonical relation appearing merely as a
special case. Because dynamics alone do not uniquely fix the quantum
algebra, canonical quantization is not the only mathematically consistent
scheme. This finding laid the groundwork for Wigner quantization, which
adopts the equations of motion as its fundamental starting point.

In 1951, Yang developed an explicit realization of Wigner's generalized
algebra \cite{Yang}. By imposing specific conditions on the Hilbert space,
he introduced a generalized momentum operator that incorporates the
reflection (parity) operator:%
\begin{equation*}
R\psi \left( x\right) =\psi \left( -x\right) .
\end{equation*}

This reflection operator modifies the momentum operator while preserving the
underlying Heisenberg equations of motion. Two years later, Green expanded
this approach to a broader class of commutation and anticommutation
relations \cite{Green}. This led to the development of parastatistics, a
framework naturally described by the representation theory of compact and
non-compact Lie algebras.

Yang's generalized momentum operator is now known to be mathematically
equivalent to the one-dimensional Dunkl operator introduced by Charles Dunkl
in the study of differential-difference operators associated with finite
reflection groups \cite{Dunkl}. This equivalence establishes a direct
connection between Wigner quantization and Dunkl analysis. Within this
framework, the reflection operator naturally appears in the quantum
dynamics, and many properties of the Wigner oscillator can be understood
using the theory of Dunkl operators. This connection has motivated extensive
research on generalized harmonic oscillators \cite{VincentX,GenestVX,Zhedanov}, Calogero type models \cite{Vinet}, special functions \cite{Orthogonal}, supersymmetric quantum mechanics \cite{dong},  and quantum systems with reflection symmetry  \cite{Isaac2016,
Jang2016,Salazar2017,Salazar2018,Mota20181,Sargol2018,Hassanabadi2019,
Sami2019,Ghazouani2020,Ojeda2020,Chung2020,Chung2021,Mota20211,
Mota20212,Bilel20221,Merad2021,Merad2022, Dong2021,
Bilel20222,Bilel20223,Halberg2022,Samira2022,Dong2023,
Junker2023, Samira2023,
Fateh20231,Fateh20232,Quesne2023,Quesne2024,Mota20241,Mota20242,Schulze20242,Junker2024,Schulze20241,Schulze20243,Ballesteros2024,bilelscr,bilelmpla}. Furthermore, the Dunkl operator has been extended in several multi-parameter \cite{Huang}. Such extensions have given rise to new classes of operators that have found applications in specific Calogero Sutherland models as well as in various problems in physics  \cite{Huang,Mota20221,Rouabhia2023,GJunker}.

On the other side, the idea of the energy dependent potentials have been studied for many years in the framework of exactly solvable wave equations \cite{Pauli,Snyder}. They are important in many areas of physics and appear naturally in different physical systems. Examples include perturbations of black holes and neutron stars \cite{Kokkotas,Nollert,Berti,Konoplya}, analog gravity \cite{Barcelo}, and relativistic and non-relativistic quantum systems in nuclear, particle, and atomic physics \cite{Rizov,Sazdjian,Mourad}. They are also applied to the study of soliton propagation \cite{Mollenauer,Serkin}. The introduction of energy dependent potentials into wave equations leads to several important consequences. One of the main effects is the modification of the scalar product to preserve the norm of the wave function \cite{Lepage,Sazd,Formanek}. Although these models extend the standard formulation of quantum mechanics, they provide an effective phenomenological framework for describing various physical systems. Energy dependent confining potentials have been successfully applied to heavy quark systems \cite{Lombard} and to bound systems of N bosons \cite{Mares}. In addition, both relativistic and nonrelativistic quantum systems with energy-dependent interactions have been investigated using different analytical approaches, including supersymmetric quantum mechanics \cite{Yekken}, Darboux transformations \cite{Schulze}, exceptional orthogonal polynomials \cite{Roy}, path integral techniques \cite{Benchikha,chetouani,meradamar}, and exact solution methods \cite{Langueur}.

Motivated by the success of both Dunkl quantum mechanics and energy-dependent potentials, we investigate their interplay within a unified framework. The incorporation of energy-dependent interactions into the Dunkl--Schrödinger equation enriches its mathematical structure while preserving the possibility of obtaining exact analytical solutions and enriching the associated spectral theory. Such an extension gives rise to a new class of exactly solvable models characterized by reflection symmetry and energy dependent interactions, opening new avenues for the study of special functions, spectral properties, and quantum systems with generalized symmetries.  An important contribution in this direction was made by Schulze-Halberg and Roy, who developed Darboux transformations for Dunkl--Schrödinger equations with energy dependent potential and position dependent mass \cite{Pinaki}.

The main objective of this work is to develop a generalized Dunkl quantum framework incorporating energy-dependent interactions through a two-parameter extension of the Dunkl differential operator. Within this formalism, we derive the modified continuity equation, probability current, and normalization condition required for a consistent probabilistic interpretation. As an application, we investigate an exactly solvable harmonic oscillator with a linearly energy-dependent potential, obtain exact analytical expressions for its energy spectrum and eigenfunctions, and analyze how the generalized Dunkl parameters and the energy dependence affect the spectral and thermodynamic properties of the system. This paper is organized as follows. In Section \ref{sec2}, we introduce the generalized Dunkl differential operator and formulate the Dunkl–Schr\"{o}dinger equation with an energy-dependent potential. We derive the corresponding continuity equation, conserved probability current, and modified normalization condition. In Section \ref{sec3}, we study an exactly solvable harmonic oscillator with a linearly energy-dependent interaction, obtaining closed-form expressions for the eigenfunctions and energy spectrum, followed by a discussion of the effects of the model parameters on the spectral properties. Section \ref{sec4} is devoted to the thermodynamic analysis, where the partition function is evaluated using the Euler–Maclaurin summation formula and the entropy, internal energy, and specific heat are investigated. Finally, Section \ref{sec5} summarizes the principal results and discusses possible directions for future research.
\section{Equation of Continuity and Modified Norm}
\label{sec2}
Within the framework of Dunkl quantum mechanics, the conventional momentum
operator is replaced by the Dunkl momentum operator,%
\begin{equation}
p\rightarrow \frac{\hbar }{i}D=\frac{\hbar }{i}\left[ \frac{d}{dx}+\frac{\nu 
}{x}\left( 1-R\right) \right] ,
\end{equation}

where $\nu $ denotes the Wigner parameter. It should be noted that, in this
expression, the reflection operator $R$ acts only on the second term, while
its contribution to the derivative term is not included. To incorporate this
missing contribution, we introduce a generalized Dunkl differential operator
depending on two parameters [],%
\begin{equation}
\mathcal{D}=\frac{d}{dx}+\frac{\nu }{x}(1-R)+\gamma \frac{d}{dx}R.
\end{equation}%
In this expression, $\nu$ is the Wigner parameter controls the strength of the reflection contribution through, while $\gamma$ is a dimensionless deformation parameter that measures the strength of the derivative-reflection coupling. The generalized operator reduces to the standard Dunkl derivative for $\gamma=0$, awhile the condition $-1<\gamma <1$  is assumed throughout this work and justified later.
The introduction of this generalized operator modifies the Heisenberg
algebra, yielding the commutation relation

\begin{equation}
\left[ x,\frac{\hbar }{i}\mathcal{D}\right] =i\hbar \left( 1+2\nu -\gamma
-2\gamma x\frac{\partial }{\partial x}R\right) .
\end{equation}%
Furthermore, the square of the generalized Dunkl operator can be written as%
\begin{equation}
\mathcal{D}^{2}=(1-\gamma ^{2})\left[ \frac{d^{2}}{dx^{2}}+\frac{2\eta }{x}%
\frac{d}{dx}-\frac{\eta }{x^{2}}(1-R)\right] ,
\end{equation}

where $\eta =\frac{\nu }{1-\gamma }$. Using this generalized derivative, the
time-dependent Schr\"{o}dinger equation with an energy-dependent potential
is then written as

\begin{equation}
-\frac{\hbar ^{2}}{2m}\mathcal{D}^{2}\psi \left( x,t\right) +V\left(
x,i\hbar \frac{\partial }{\partial t}\right) \psi \left( x,t\right) =i\hbar 
\frac{\partial }{\partial t}\psi \left( x,t\right) ,  \label{Sch1}
\end{equation}%
while its complex conjugate takes the form

\begin{equation}
-\frac{\hbar ^{2}}{2m}\mathcal{D}^{2}\psi ^{\ast }\left( x,t\right) +\left[
V\left( x,i\hbar \frac{\partial }{\partial t}\right) \psi \left( x,t\right) %
\right] ^{\ast }=-i\hbar \frac{\partial }{\partial t}\psi ^{\ast }\left(
x,t\right) .  \label{Sch2}
\end{equation}

The purpose of this section is to derive the continuity equation and the
associated normalization condition for the generalized Dunkl system.
Multiplying Eq.(\ref{Sch1}) by $\psi ^{\ast }$ and Eq.(\ref{Sch2}) by $\psi $%
, followed by subtracting the resulting expressions, yields

\begin{equation}
i\hbar \left( \psi ^{\ast }\frac{\partial }{\partial t}\psi +\psi \frac{%
\partial }{\partial t}\psi ^{\ast }\right) =-\frac{\hbar ^{2}}{2m}\left(
\psi ^{\ast }\mathcal{D}^{2}\psi -\psi \mathcal{D}^{2}\psi ^{\ast }\right) +%
\left[ \psi ^{\ast }V(x,i\hbar \frac{\partial }{\partial t})\psi -\psi
\lbrack V(x_{i},i\hbar \frac{\partial }{\partial t})\psi ]^{\ast }\right] .
\label{sch}
\end{equation}

Since the Hamiltonian is independent of time, the wave function can be
expressed in terms of stationary states,%
\begin{equation}
\psi \left( x,t\right) =\sum_{n}a_{n}\psi _{n}\left( x,t\right)
=\sum_{n}a_{n}e^{-\frac{iE_{n}}{\hbar }t}\Psi _{n}\left( x\right) .
\end{equation}

The generalized Dunkl operator is Hermitian with respect to the weighted
inner product defined by the Dunkl measure $\left\vert x\right\vert ^{2\eta }
$. Multiplying Eq.(\ref{sch}) by this weight and employing the explicit form
of $\mathcal{D}^{2}$, the spatial terms can be arranged into a total
derivative. As a result, the continuity equation takes the form 
\begin{equation}
\frac{\partial }{\partial t}\rho +\frac{d}{dx}J=f\left( x\right) ,
\end{equation}

where 
\begin{equation}
\rho =\sum_{n,k}a_{k}^{\ast }a_{n}\left\vert x\right\vert ^{2\eta }\psi
_{k}^{\ast }\left[ 1-\frac{V\left( x,E_{k}\right) -V\left( x,E_{n}\right) }{%
E_{k}-E_{n}}\right] \psi _{n},
\end{equation}%
\begin{equation}
J=\sum_{n,k}a_{k}^{\ast }a_{n}\frac{\hbar (1-\gamma ^{2})}{2im}\left\vert
x\right\vert ^{2\eta }\left[ \psi _{k}^{\ast }\frac{d\psi _{n}}{dx}-\psi _{n}%
\frac{d\psi _{k}^{\ast }}{dx}\right] ,
\end{equation}

and \bigskip 
\begin{equation}
f\left( x\right) =\sum_{n,k}a_{k}^{\ast }a_{n}\frac{i\hbar (1-\gamma
^{2})\left\vert x\right\vert ^{2\eta }}{2mx^{2}}\left[ \psi _{k}^{\ast
}R\psi _{n}-\psi _{n}R\psi _{k}^{\ast }\right] .
\end{equation}%
The function $f\left( x\right) $ originates from the reflection part of the
generalized Dunkl operator and acts as a source term in the continuity
equation. This contribution can be incorporated into the definition of the
probability current by introducing the modified current 
\begin{equation}
\mathcal{J}=\sum_{n,k}a_{k}^{\ast }a_{n}\left\{ \frac{\hbar (1-\gamma ^{2})}{%
2im}\left\vert x\right\vert ^{2\eta }\left[ \psi _{k}^{\ast }\frac{d\psi _{n}%
}{dx}-\psi _{n}\frac{d\psi _{k}^{\ast }}{dx}\right] +\int^{x}\frac{i\hbar
(1-\gamma ^{2})\left\vert y\right\vert ^{2\eta }}{2my^{2}}\left[ \psi
_{k}^{\ast }R\psi _{n}-\psi _{n}R\psi _{k}^{\ast }\right] dy\right\} .
\end{equation}%
The continuity equation then assumes the standard conservation form,%
\begin{equation}
\frac{\partial }{\partial t}\rho +\frac{d}{dx}\mathcal{J}=0.
\end{equation}%
The orthogonality relation follows by integrating the probability density over the entire configuration space. One obtains
\begin{equation}
\left\langle \psi _{m}\right. \left\vert \psi _{n}\right\rangle =\int
dx\left\vert x\right\vert ^{2\eta }\psi _{m}^{\ast }\left[ 1-\frac{V\left(
x,E_{m}\right) -V\left( x,E_{n}\right) }{E_{m}-E_{n}}\right] \psi
_{n}=\delta _{nm}.
\end{equation}

Taking the limit $n\rightarrow m$ yields the modified normalization
condition,%
\begin{equation}
\int dx\left\vert x\right\vert ^{2\eta }\psi _{n}^{\ast }\left[ 1-\frac{%
\partial V\left( x,E_{n}\right) }{\partial E_{n}}\right] \psi _{n}=1.\label{nor}
\end{equation}%
Therefore, the probability density in the presence of an energy-dependent
potential and the generalized Dunkl derivative differs from the conventional
expression. The wave function must satisfy the modified normalization
condition above, and the quantity inside the integral must remain
non-negative to ensure a consistent probabilistic interpretation.

\section{ Harmonic Oscillator Potential with Linear \texorpdfstring{$E$}{E}-dependence}
\label{sec3}
To illustrate the formalism, we consider an exactly solvable model
describing a harmonic oscillator with an energy-dependent frequency. The
one-dimensional Schr\"{o}dinger equation involving the generalized Dunkl
operator is given by

\begin{equation}
\left\{ -\frac{\hbar ^{2}}{2m}\left( 1-\gamma ^{2}\right) \left[ \frac{d^{2}%
}{dx^{2}}+\frac{2\eta }{x}\frac{d}{dx}-\frac{\eta }{x^{2}}(1-R)\right]
+V\left( x,E\right) \right\} \Psi \left( x\right) =E\Psi \left( x\right) ,
\label{Sche}
\end{equation}

where the potential is chosen as

\begin{equation}
V\left( x,E\right) =\frac{m\omega ^{2}}{2}\left( 1+\lambda E\right) x^{2},\label{fun2}
\end{equation}

and $\lambda $ denoting the energy-dependence parameter. Such a linear
energy dependence has been shown to provide a consistent quantum-mechanical
framework and can be mapped onto an equivalent formulation involving a
nonlocal potential \cite{Mares,Yekken}.

Since the Hamiltonian commutes with the reflection operator,%
\begin{equation}
\left[ H,R\right] =0,
\end{equation}

the eigenfunctions can be chosen as simultaneous eigenstates of both
operators,%
\begin{equation}
R\Psi \left( x\right) =s\Psi \left( x\right) ,\text{ \ with }s=\pm 1,
\label{par}
\end{equation}

where $s=1$ and $s=-1$ correspond to the even- and odd-parity sectors,
respectively. Consequently, the Hilbert space decomposes into two
independent parity sectors, and the Schr\"{o}dinger equation can be solved
separately for each one.

Substituting (\ref{par}) into Eq.(\ref{Sche}) leads to%
\begin{equation}
\left\{ \frac{d^{2}}{dx^{2}}+\frac{2\eta }{x}\frac{d}{dx}-\frac{\eta }{x^{2}}%
(1-s)-\frac{m^{2}\omega ^{2}}{\hbar ^{2}}\frac{\left( 1+\lambda E\right) }{%
\left( 1-\gamma ^{2}\right) }x^{2}+\frac{2mE}{\left( 1-\gamma ^{2}\right)
\hbar ^{2}}\right\} \Psi ^{s}\left( x\right) =0.  \label{simp}
\end{equation}

To transform Eq.(\ref{simp}), into a standard form, we introduce the
dimensionless variable%
\begin{equation}
z=\frac{m\omega }{\hbar }\sqrt{\frac{1+\lambda E}{1-\gamma ^{2}}}x^{2}.
\end{equation}

Eq.(\ref{simp}) then becomes 
\begin{equation}
\left\{ z\frac{d^{2}}{dz^{2}}+\left( \eta +\frac{1}{2}\right) \frac{d}{dz}-%
\frac{\eta }{4z}(1-s)-\frac{z}{4}+\frac{E}{2\hbar \omega \sqrt{\left(
1+\lambda E\right) \left( 1-\gamma ^{2}\right) }}\right\} \Psi ^{s}\left(
z\right) =0.  \label{solve}
\end{equation}%
The parity parameter "$s$" is retained explicitly throughout the derivation,
allowing the even ($s=1$) and odd ($s=-1$) solutions to be obtained within a
unified analytical framework. To determine the normalizable solutions, we
factor out the asymptotic behavior by introducing

\begin{equation}
\Psi ^{s}\left( z\right) =z^{\frac{1-s}{4}}e^{-\frac{z}{2}}\Phi ^{s}\left(
z\right) .  \label{Ana}
\end{equation}%
Substituting Eq.~(\ref{Ana}) into Eq.~(\ref{solve}) gives

\begin{equation}
\left\{ z\frac{d^{2}}{dz^{2}}+\left( \eta -\frac{s}{2}+1-z\right) \frac{d}{dz%
}-\frac{1}{2}\left( \eta +1-\frac{s}{2}\right) +\frac{E}{2\hbar \omega \sqrt{%
\left( 1+\lambda E\right) \left( 1-\gamma ^{2}\right) }}\right\} \Phi
^{s}\left( z\right) =0.  \label{Laguerre}
\end{equation}%
Eq.~(\ref{Laguerre}) is recognized as the differential equation satisfied by
the associated Laguerre polynomials. Therefore, the regular polynomial
solutions are%
\begin{equation}
\Phi ^{s}\left( z\right) =L_{n}^{\eta -\frac{s}{2}}\left( z\right) ,
\end{equation}%
provided that%
\begin{equation}
n=\frac{E}{2\hbar \omega \sqrt{\left( 1+\lambda E\right) \left( 1-\gamma
^{2}\right) }}-\frac{1}{2}\left( \eta +1-\frac{s}{2}\right) .
\end{equation}%
$n=0,1,2,\ldots$. This quantization condition determines the energy spectrum as%

\begin{eqnarray}
E_{n}^{s} &=&2\hbar \omega \left( n+\frac{\eta +1-\frac{s}{2}}{2}\right) \Bigg[
\left( 1-\gamma ^{2}\right) \lambda \hbar \omega \left( n+\frac{\eta +1-\frac{s}{2}}{2}\right) \nonumber \\
&&+ \sqrt{\left( 1-\gamma ^{2}\right) + \left( 1-\gamma ^{2}\right)^{2} \lambda ^{2} \hbar ^{2} \omega ^{2} \left( n+\frac{\eta +1-\frac{s}{2}}{2} \right)^{2}} \Bigg].\label{spec}
\end{eqnarray}

The energy spectrum exhibits several distinctive features arising from the
combined effects of the generalized Dunkl operator and the energy-dependent
interaction. First, the eigenvalues depend explicitly on the parity quantum
number $s=\pm 1$, indicating that the reflection symmetry removes the
degeneracy between the even and odd parity sectors. This parity splitting is
a characteristic consequence of the Dunkl formalism and is absent in the
ordinary one dimensional harmonic oscillator. The Wigner parameter $\nu $
shifting the energy levels and modifying the separation between the parity
branches. In contrast, the deformation parameter $\gamma $ influences both
the kinetic and potential contributions. Indeed, it appears through the
factor $\left( 1-\gamma ^{2}\right) $, which renormalizes the effective
oscillator frequency, while simultaneously modifying the parameter $\eta $.
Consequently, $\gamma $ controls both the overall energy scale and the
parity splitting. The restriction $\left\vert \gamma \right\vert <1$ ensures
that the effective kinetic energy operator remains positive definite and
guarantees the reality of the energy spectrum.

Unlike the equally spaced spectrum of the conventional harmonic oscillator,
the present model exhibits a nonlinear dependence on the principal quantum
number. Introducing%
\begin{equation}
N=n+\frac{\eta +1-\frac{s}{2}}{2},
\end{equation}%
the asymptotic expansion of the exact spectrum for large $n$, showing that
the leading contribution is quadratic,%
\begin{equation}
E_{n}^{s}\propto n^{2},
\end{equation}%
Consequently, the spacing between successive energy levels increases with
the quantum number,%
\begin{equation}
\Delta E_{n}^{s}\propto n,
\end{equation}%
instead of remaining constant as in the ordinary oscillator. This
progressive broadening of the spectral lines is a direct consequence of the
energy dependent potential and constitutes one of the principal signatures
of the present model.

The eigenstates associated with the energy-dependent harmonic oscillator in
the generalized Dunkl formalism are given by%
\begin{equation}
\Psi ^{s}\left( x\right) =\mathcal{N}x^{\frac{1-s}{2}}\exp \left( -\frac{%
m\omega }{2\hbar }\sqrt{\frac{1+\lambda E_{n}^{s}}{1-\gamma ^{2}}}%
x^{2}\right) L_{n}^{\eta -\frac{s}{2}}\left( \frac{m\omega }{\hbar }\sqrt{%
\frac{1+\lambda E_{n}^{s}}{1-\gamma ^{2}}}x^{2}\right) .\label{fun1}
\end{equation}%
Here $\mathcal{N}$ represents the normalization constant, which is
determined from the modified normalization condition. By substituting Eqs.
(\ref{fun2}) and (\ref{fun1}) into Eq. (\ref{nor}) and employing the orthogonality relation%
\begin{equation}
\int_{0}^{+\infty }x^{\alpha }e^{-x}L_{n}^{\eta -\frac{1}{2}}\left( x\right)
L_{m}^{\eta -\frac{1}{2}}\left( x\right) dx=\frac{\Gamma \left( n+\alpha
+1\right) }{n!}\delta _{nm}.
\end{equation}%
the normalized wavefunctions take the form%
\begin{eqnarray}
\Psi _{n}^{s}\left( x\right)  &=&\sqrt{\frac{n!\left( \frac{m\omega }{\hbar }%
\sqrt{\frac{1+\lambda E_{n}^{s}}{1-\gamma ^{2}}}\right) ^{\eta +1-\frac{s}{2}%
}}{\left( 1-\frac{\lambda \hbar \omega }{2}\sqrt{\frac{1-\gamma ^{2}}{%
1+\lambda E_{n}^{s}}}\left( 2n+\eta -\frac{s}{2}+1\right) \right) \Gamma
\left( n+\eta +1-\frac{s}{2}\right) }}\times   \notag \\
&&x^{\frac{1-s}{2}}e^{-\frac{m\omega }{2\hbar }\sqrt{\frac{1+\lambda
E_{n}^{s}}{1-\gamma ^{2}}}x^{2}}L_{n}^{\eta -\frac{s}{2}}\left( \frac{%
m\omega }{\hbar }\sqrt{\frac{1+\lambda E_{n}^{s}}{1-\gamma ^{2}}}%
x^{2}\right) .
\end{eqnarray}%
As follows from Eq. (\ref{spec}), the complete spectrum, incorporating both even-
and odd-parity solutions, is no longer uniformly spaced. This behavior
differs markedly from that of the conventional singular harmonic oscillator,
whose energy levels are equally separated. To emphasize the underlying
parity structure, the spectrum can be rewritten in the compact form

\begin{equation}
E_{n}^{s}=\frac{\hbar \omega }{2}\left( k+\eta +\frac{1}{2}\right) \left[
\left( 1-\gamma ^{2}\right) \lambda \hbar \omega \left( k+\eta +\frac{1}{2}%
\right) +\sqrt{4\left( 1-\gamma ^{2}\right) +\left( 1-\gamma ^{2}\right)
^{2}\lambda ^{2}\hbar ^{2}\omega ^{2}\left( k+\eta +\frac{1}{2}\right) ^{2}}%
\right] ,
\end{equation}

where 
\begin{equation}
p=\frac{1-s}{2}\text{ and }k=2n+p.
\end{equation}

Several limiting situations can be identified directly from the above
expression.

\begin{figure}[H]
\begin{subfigure}[t]{0.45\textwidth}
        \centering
        \includegraphics[width=0.85\linewidth]{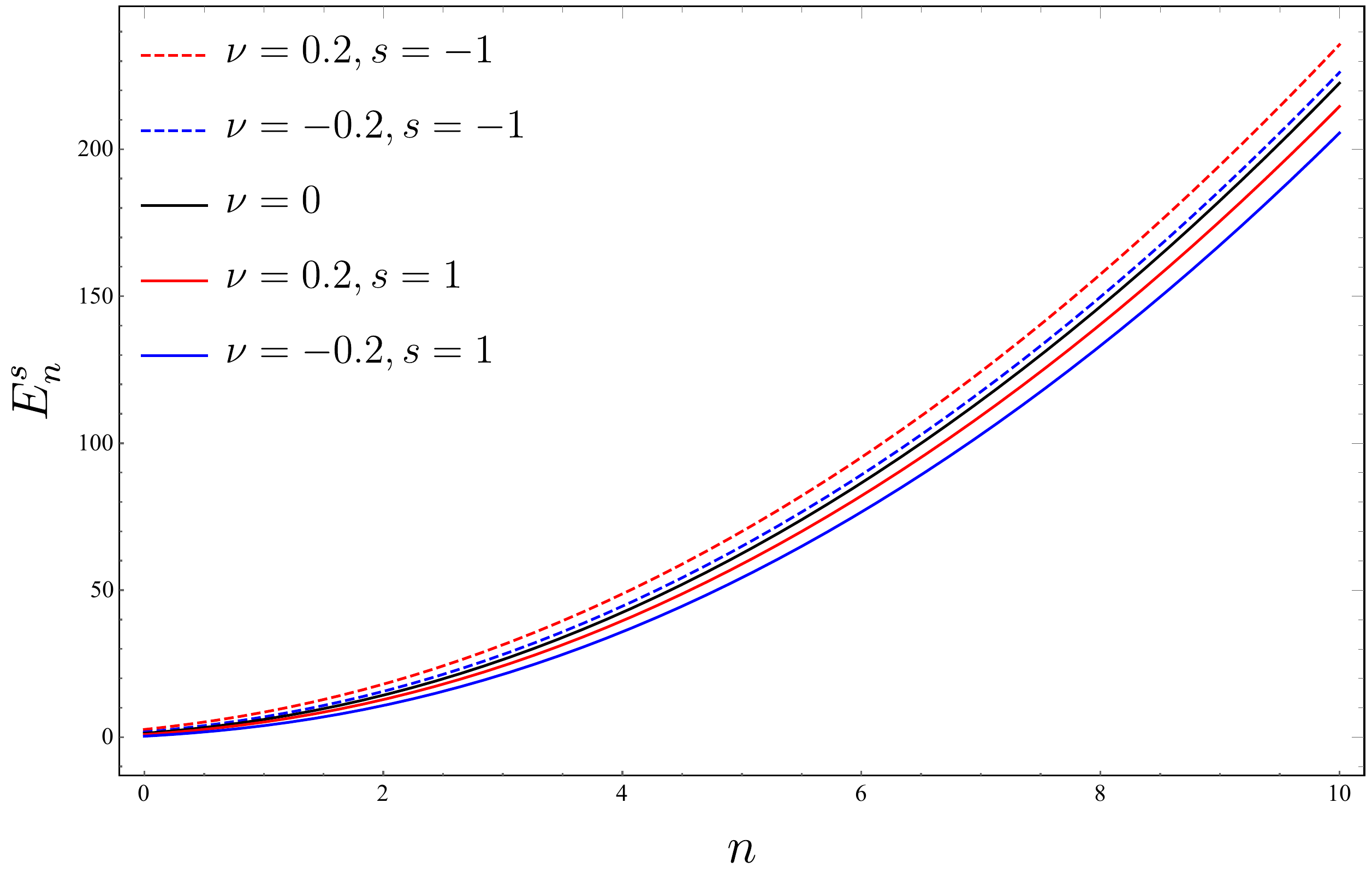}
        \subcaption{$\gamma=0.1$, $\lambda=0.5$}
       \label{fig:s1}
   \end{subfigure}\hfill
\begin{subfigure}[t]{0.45\textwidth}
        \centering
       \includegraphics[width=0.85\linewidth]{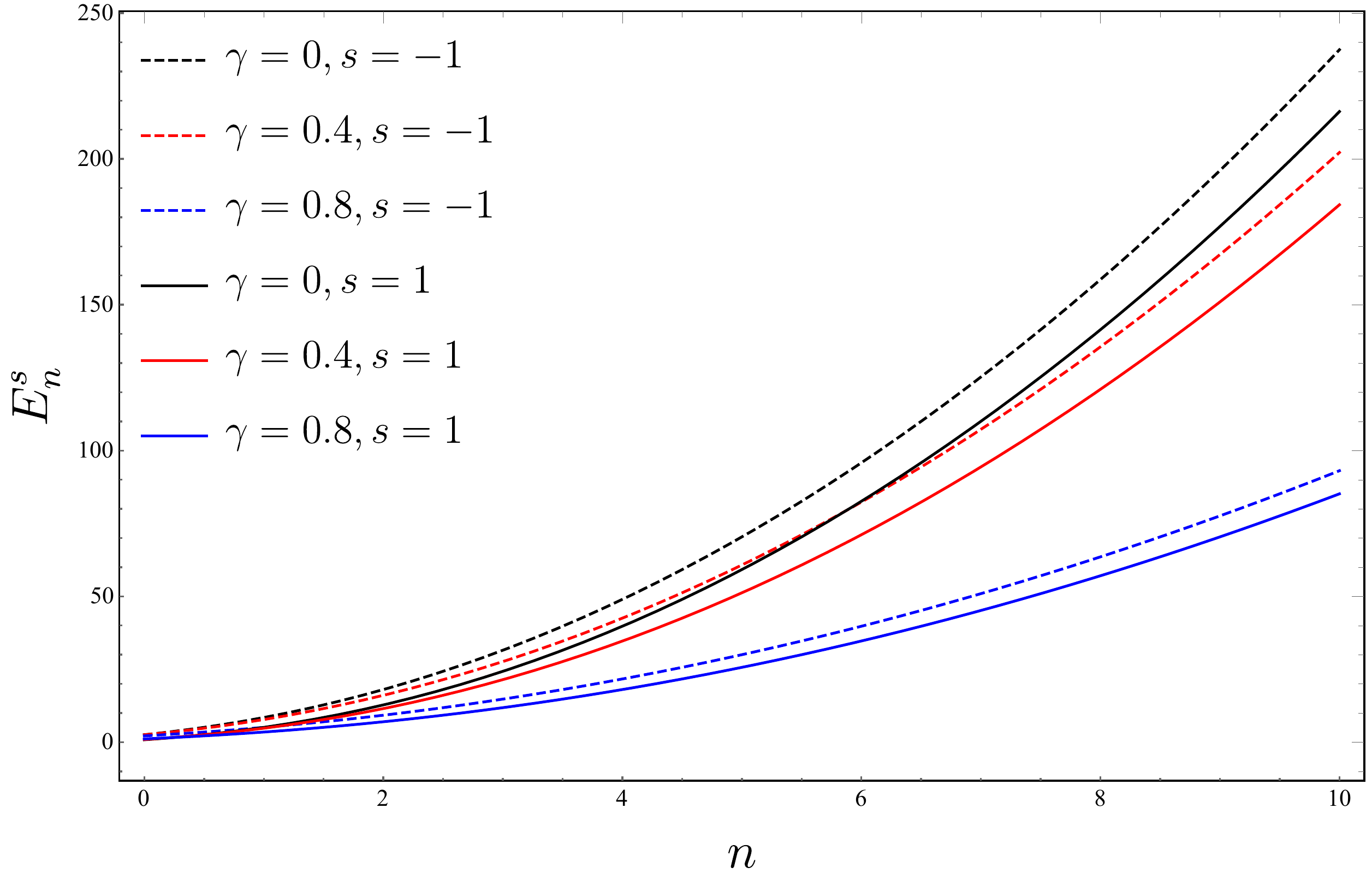}
       \subcaption{$\nu=0.2$, $\lambda=0.5$}
       \label{fig:s2}
    \end{subfigure}\hfill
    \centering
\begin{subfigure}[t]{0.45\textwidth}
     \centering
       \includegraphics[width=0.85\linewidth]{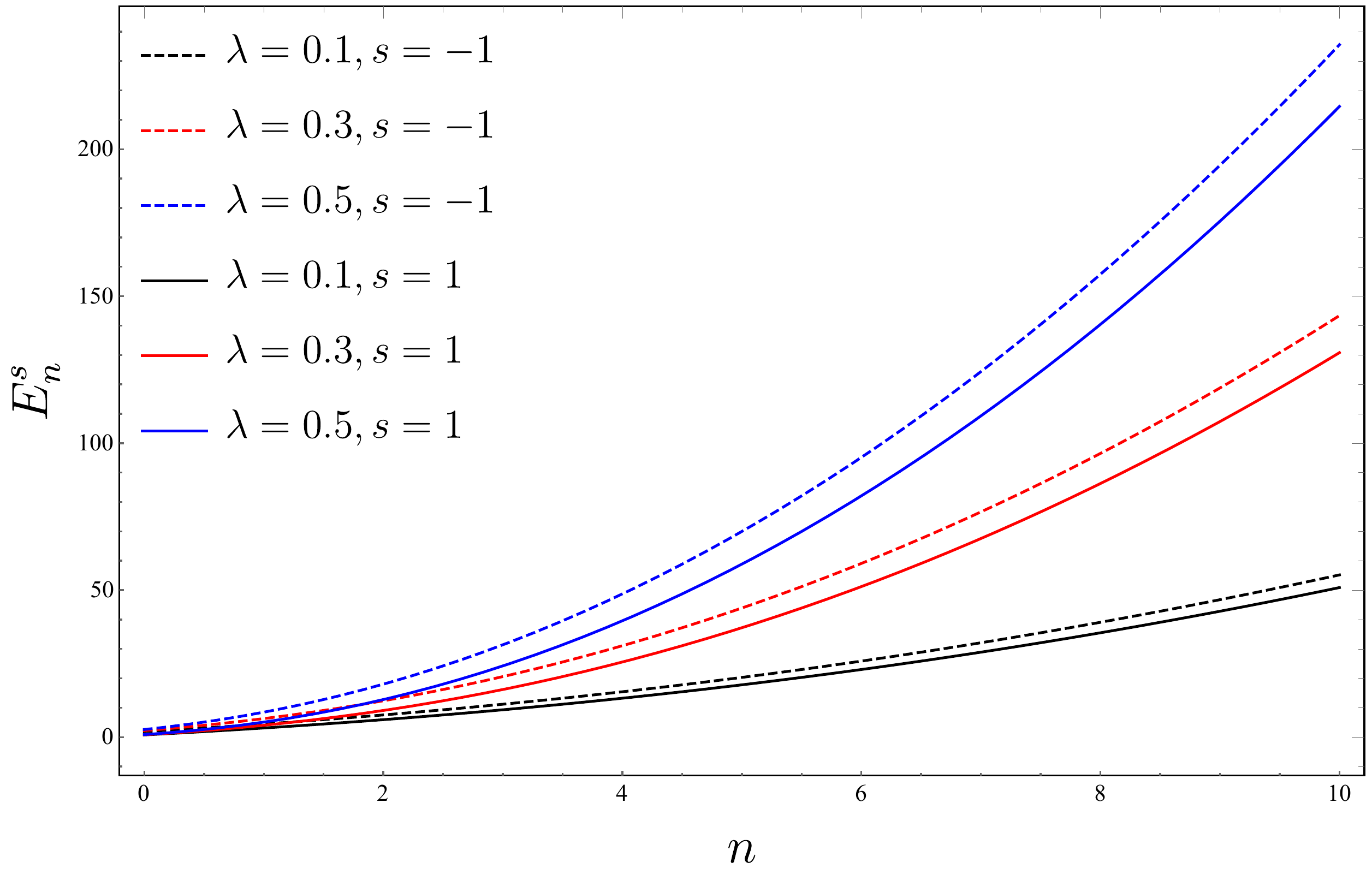}
       \subcaption{$\gamma=0.1$, $\nu=0.2$}
       \label{fig:s3}
    \end{subfigure}
\caption{ The variation of the energy spectrum with the quantum number $n$ for different values of the parameters $\nu$, $\gamma$, and $\lambda$.}
\label{fig:spectre}
\end{figure}

\begin{enumerate}
\item  $\lambda \rightarrow 0:$ In the absence of the energy dependence the
spectrum simplifies to%
\begin{equation}
E_{n}^{s}=2\hbar \omega \sqrt{1-\gamma ^{2}}\left( n+\frac{\eta +1-\frac{s}{2%
}}{2}\right) ,
\end{equation}

\item For $\eta =\gamma =0:$ Eliminating the Dunkl contribution leads to%
\begin{equation}
E_{k}^{s}=\frac{\hbar \omega }{2}\left( k+\frac{1}{2}\right) \left[ \lambda
\hbar \omega \left( k+\frac{1}{2}\right) +\sqrt{4+\lambda ^{2}\hbar
^{2}\omega ^{2}\left( k+\frac{1}{2}\right) ^{2}}\right] ,
\end{equation}

which coincides with the result obtained in \cite{Formanek}.

\item For $\eta =\gamma =\lambda =0:$ In the absence of both deformation and
energy dependence, the spectrum reduces to%
\begin{equation}
E_{k}^{s}=\hbar \omega \left( k+\frac{1}{2}\right) 
\end{equation}%
which is precisely the well-known energy spectrum of the one-dimensional
quantum harmonic oscillator.
\end{enumerate}
Figure \ref{fig:spectre} shows the variation of the energy spectrum with the quantum number $n$ for different values of the parameters $\nu$, $\gamma$, and $\lambda$.\\
Figure \ref{fig:s1} illustrates the effect of the Wigner parameter ($\nu$. The energy increases monotonically with the quantum number $n$, as expected for a bound system. For a fixed value of $n$, the energy levels are shifted toward higher values as $\nu$ increases. This behavior is consistent with the analytical expression of the spectrum, where $\nu$ enters through the parameter $\eta$, leading to an overall upward displacement of the energy levels. Moreover, the odd-parity states $s=-1$ remain above the corresponding even-parity states $s=1$, reflecting the parity splitting induced by the reflection operator.\\
The dependence of the spectrum on the deformation parameter $\gamma$ is presented in Fig.\ref{fig:s2}. Although the energy remains an increasing function of $n$, it decreases with increasing $\gamma$ for a fixed quantum number. This behavior follows directly from the factor "$1-\gamma ^2$" appearing in the exact solution, which effectively reduces the oscillator frequency as $\gamma$ increases. Consequently, the confinement becomes weaker, resulting in lower energy eigenvalues. The separation between the even- and odd-parity branches is preserved throughout the considered range of $\gamma$.\\
Figure \ref{fig:s3} displays the influence of the energy-dependence parameter $\lambda$. Increasing $\lambda$ shifts the spectrum toward higher energies, while the monotonic dependence on the quantum number is maintained. This behavior originates from the nonlinear coupling between the energy eigenvalue and the interaction potential, which becomes more pronounced for larger values of $\lambda$. As a result, the spacing between successive energy levels increases with $n$, in agreement with the asymptotic behavior of the exact spectrum. Furthermore, the parity splitting remains clearly visible, with the odd-parity branch lying systematically above the even-parity branch.\\
The numerical results agree well with the analytical energy spectrum. The Wigner parameter $\nu$ and the energy-dependence parameter $\lambda$ increase the energy levels, whereas the deformation parameter $\gamma$ decreases them by reducing the effective oscillator frequency. In all cases, the odd-parity states remain above the even-parity states, showing that the generalized Dunkl operator has a significant effect on the spectral properties of the system through the reflection symmetry.

\section{Thermal Properties}
\label{sec4}
\begin{figure}[H]
\begin{subfigure}[t]{0.45\textwidth}
        \centering
        \includegraphics[width=0.9\linewidth]{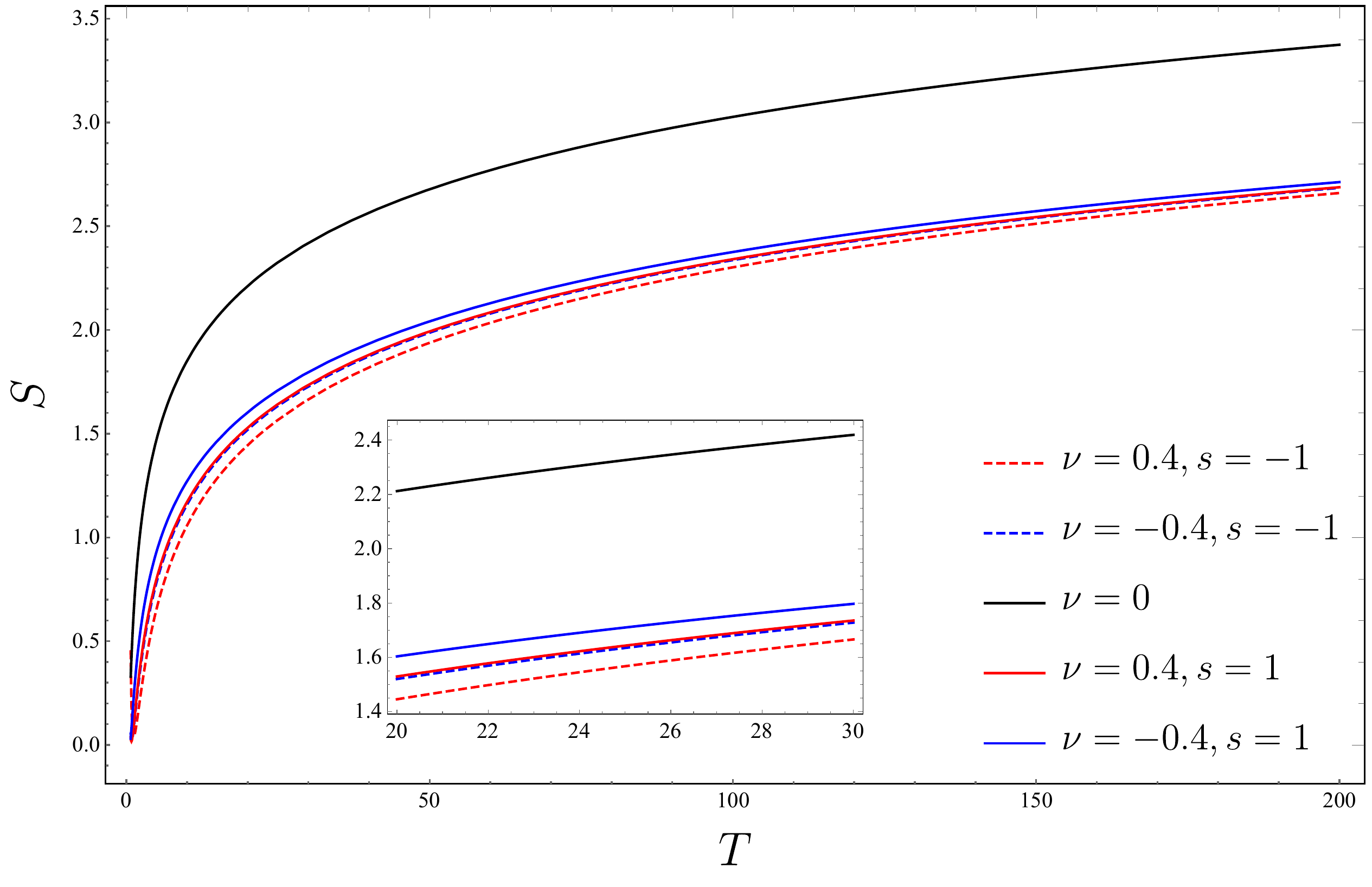}
        \subcaption{entropy}
       \label{fig:r1}
   \end{subfigure}\hfill
\begin{subfigure}[t]{0.45\textwidth}
        \centering
       \includegraphics[width=0.9\linewidth]{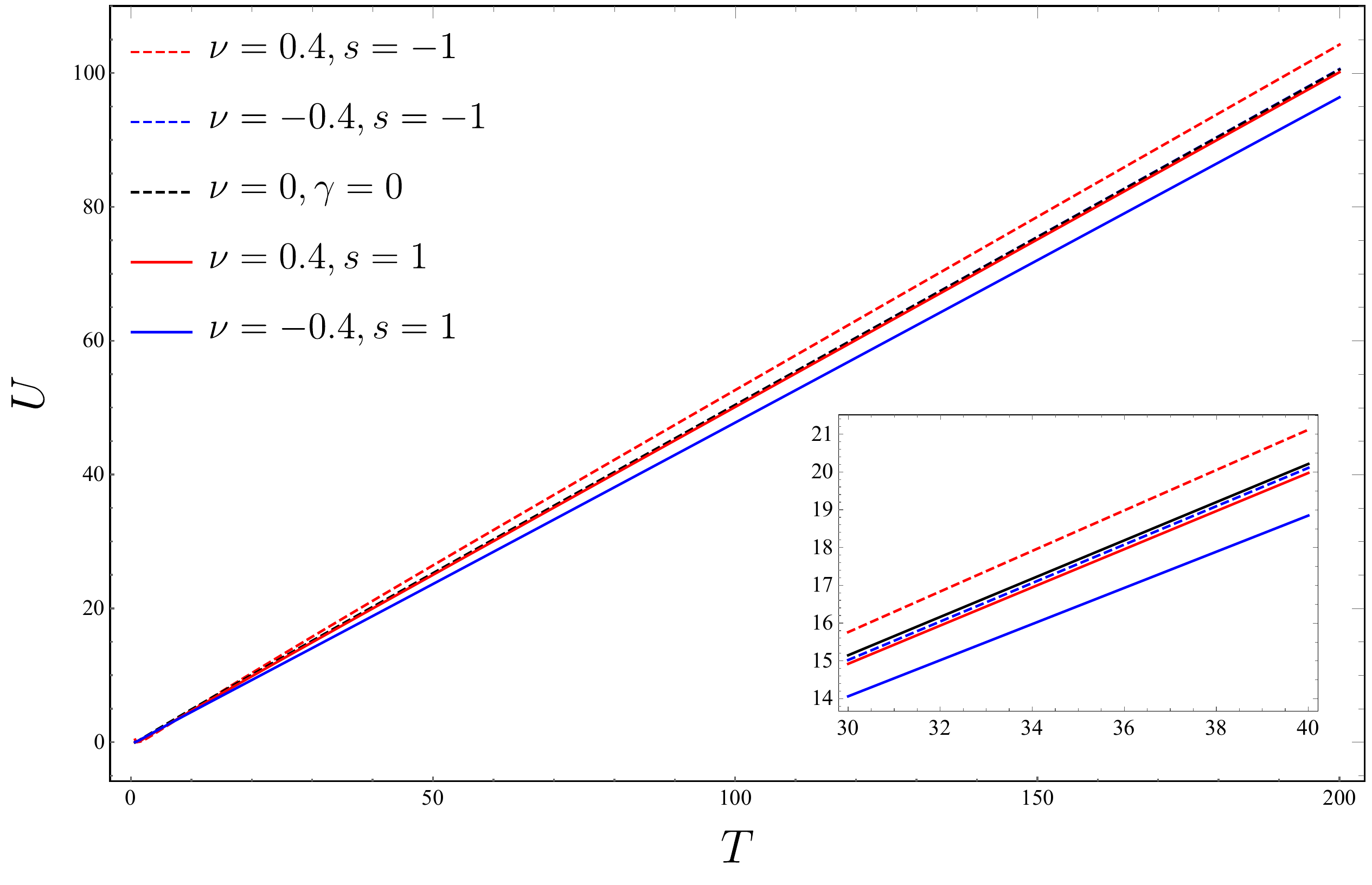}
       \subcaption{internal energy}
       \label{fig:r2}
    \end{subfigure}\hfill
    \centering
\begin{subfigure}[t]{0.45\textwidth}
     \centering
       \includegraphics[width=0.9\linewidth]{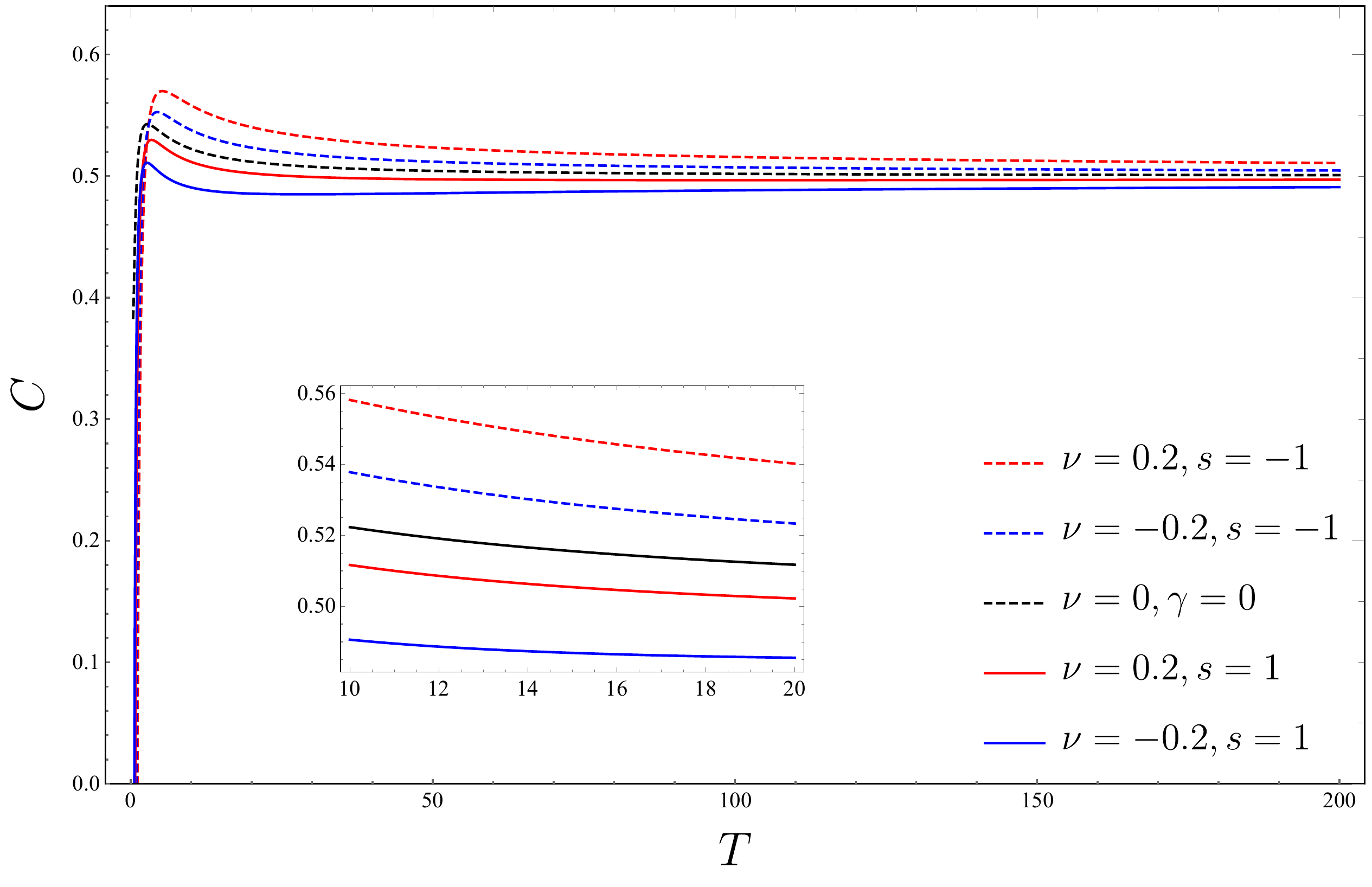}
       \subcaption{ specific heat}
       \label{fig:r3}
    \end{subfigure}
\caption{The influence of Wigner parameters $\nu$ on thermodynamic properties. Here, we use $\gamma=0.1$, $\lambda=0.5$.}
\label{fig:rho1}
\end{figure}

\begin{figure}[H]
\begin{subfigure}[t]{0.45\textwidth}
        \centering
        \includegraphics[width=0.85\linewidth]{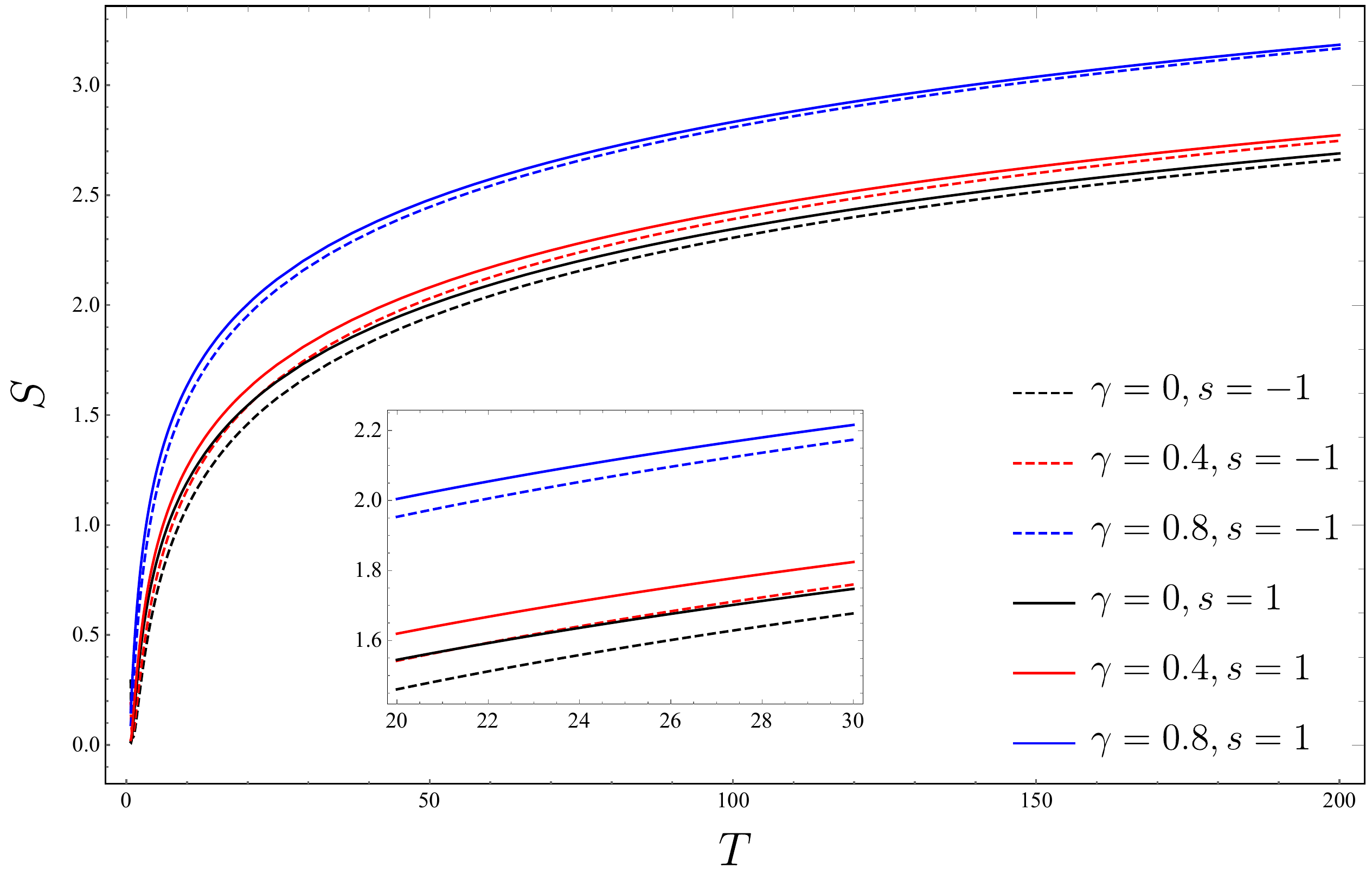}
        \subcaption{entropy}
       \label{fig:r11}
   \end{subfigure}\hfill
\begin{subfigure}[t]{0.45\textwidth}
        \centering
       \includegraphics[width=0.85\linewidth]{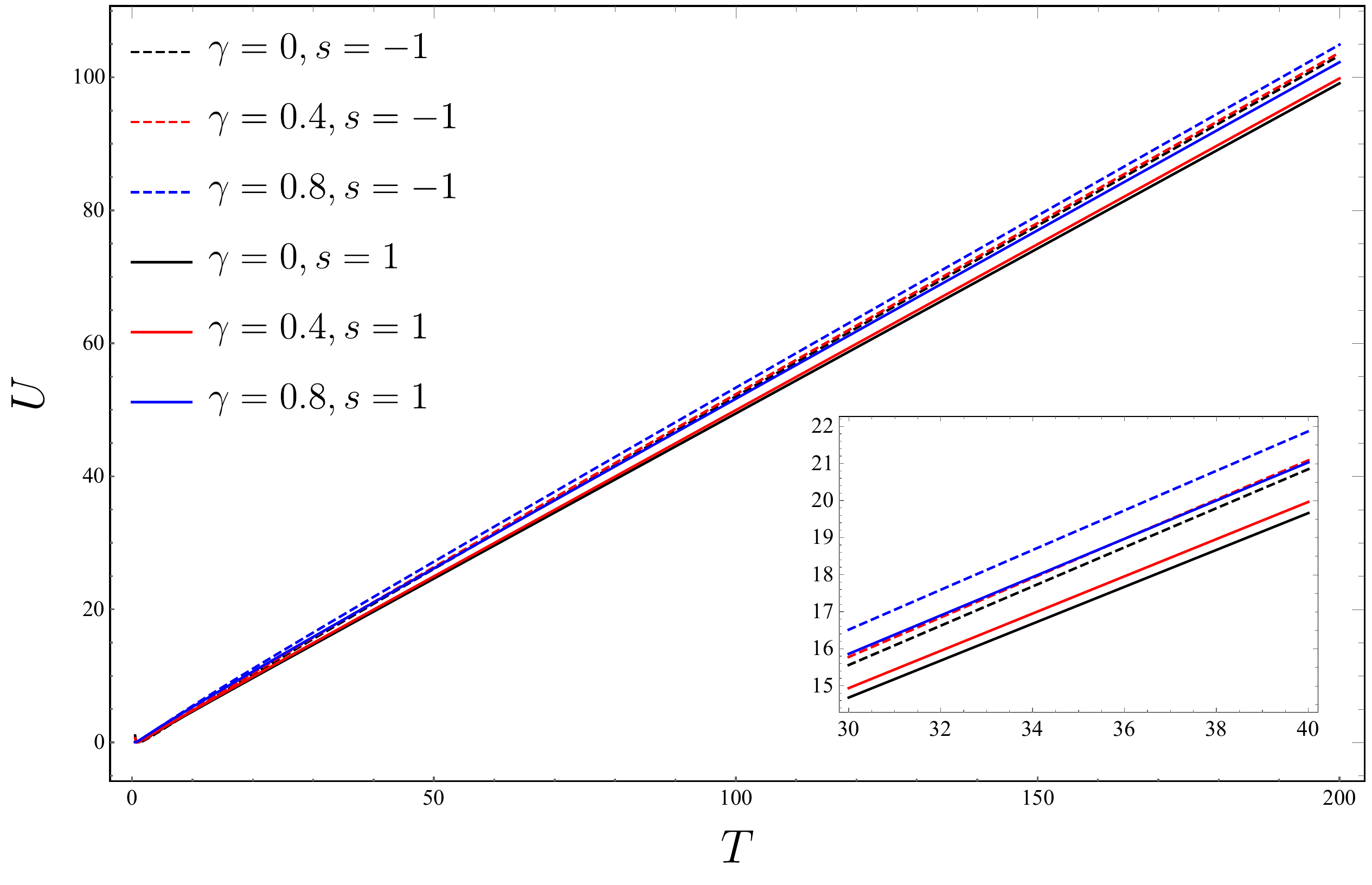}
       \subcaption{internal energy}
       \label{fig:r22}
    \end{subfigure}\hfill
    \centering
\begin{subfigure}[t]{0.45\textwidth}
     \centering
       \includegraphics[width=0.85\linewidth]{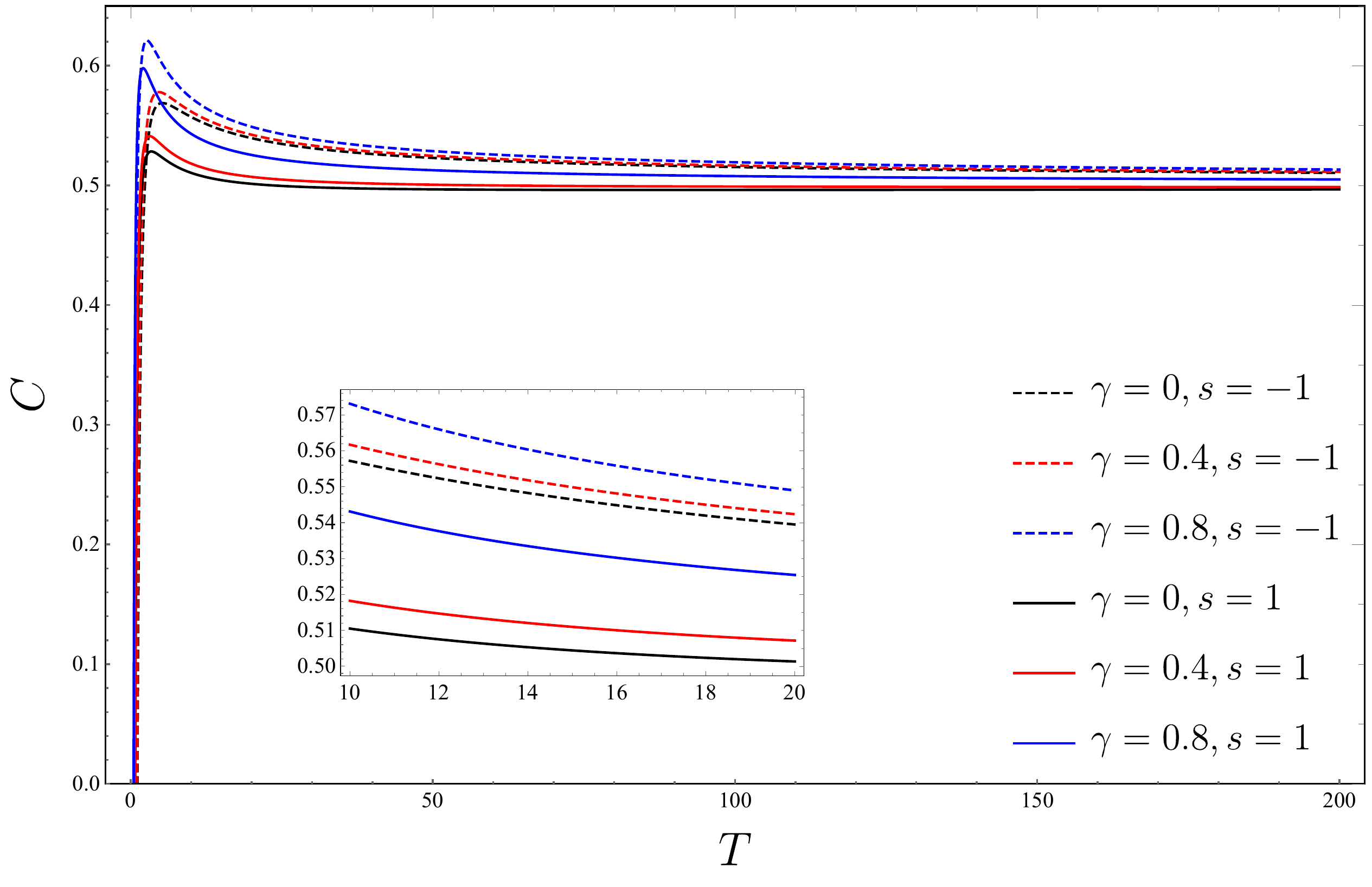}
       \subcaption{specific heat}
       \label{fig:r33}
    \end{subfigure}
\caption{ The effect of $\gamma$ on thermodynamic properties. Here, we use $\nu=0.2$, $\lambda=0.5$.}
\label{fig:rho11}
\end{figure}

\begin{figure}[H]
\begin{subfigure}[t]{0.45\textwidth}
        \centering
        \includegraphics[width=0.85\linewidth]{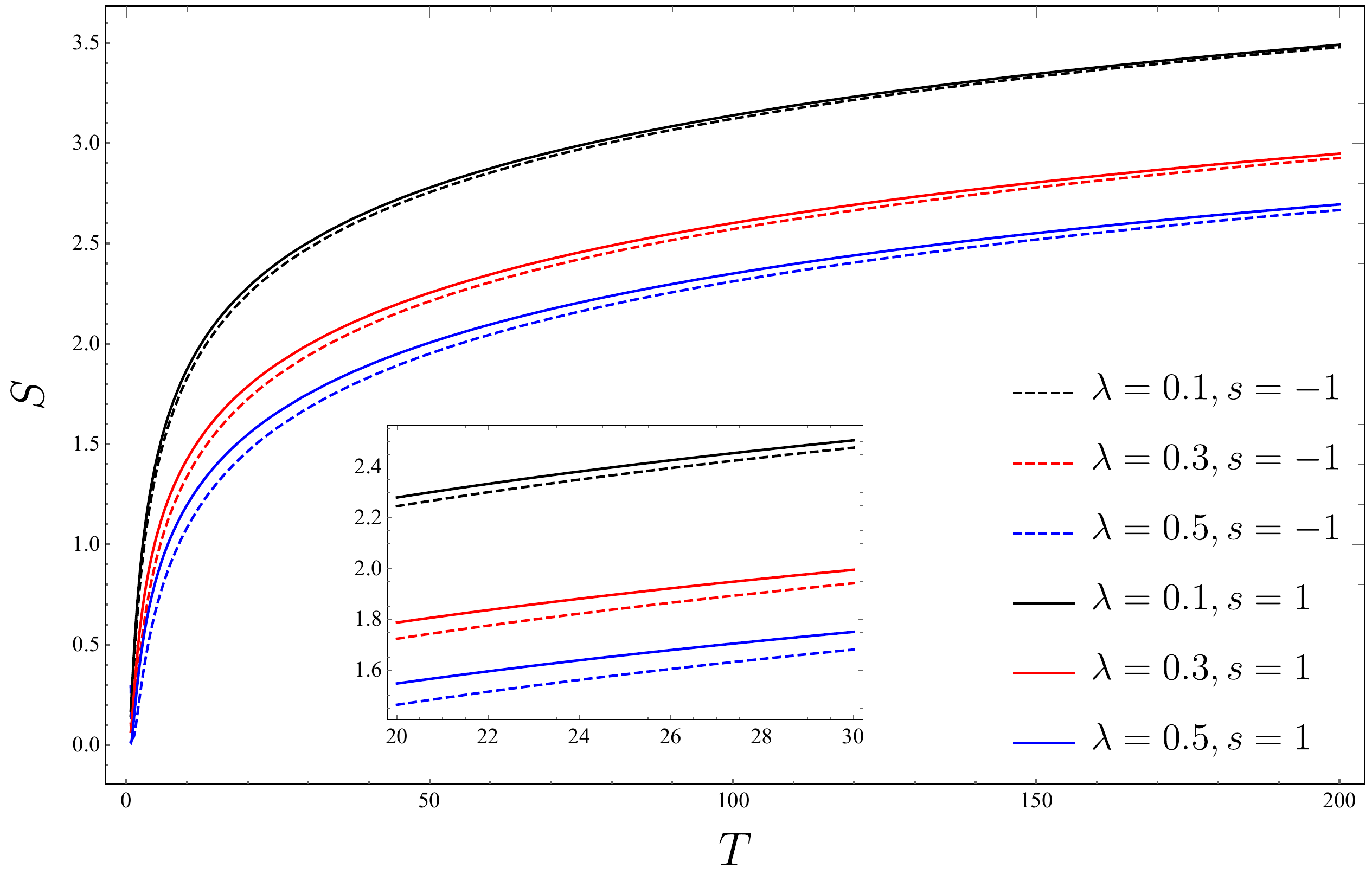}
        \subcaption{Entropy}
       \label{fig:r111}
   \end{subfigure}
   \hfill
\begin{subfigure}[t]{0.45\textwidth}
        \centering
       \includegraphics[width=0.85\linewidth]{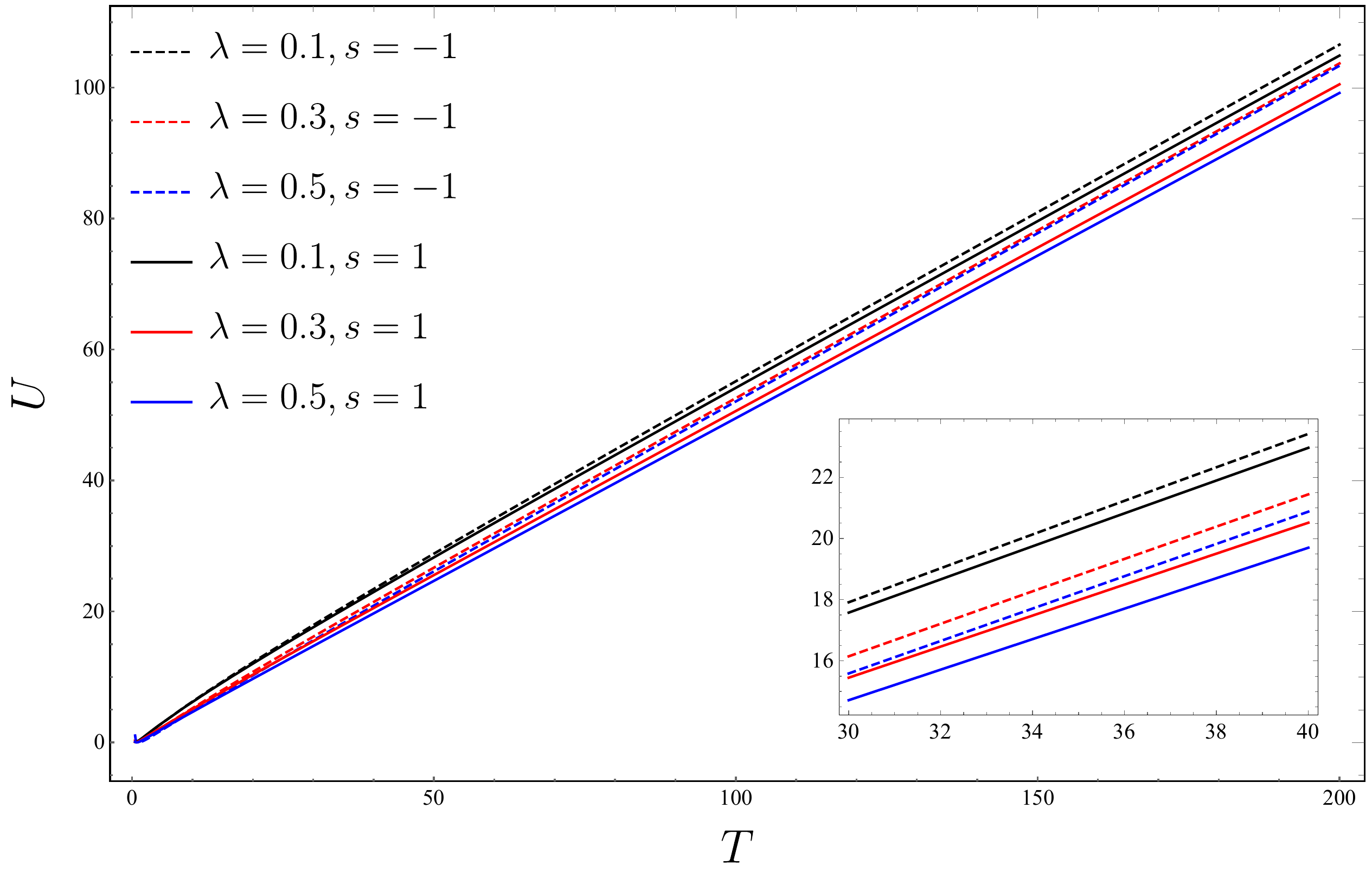}
       \subcaption{internal energy}
       \label{fig:r222}
    \end{subfigure}
    \hfill
    \centering
\begin{subfigure}[t]{0.45\textwidth}
     \centering
       \includegraphics[width=0.85\linewidth]{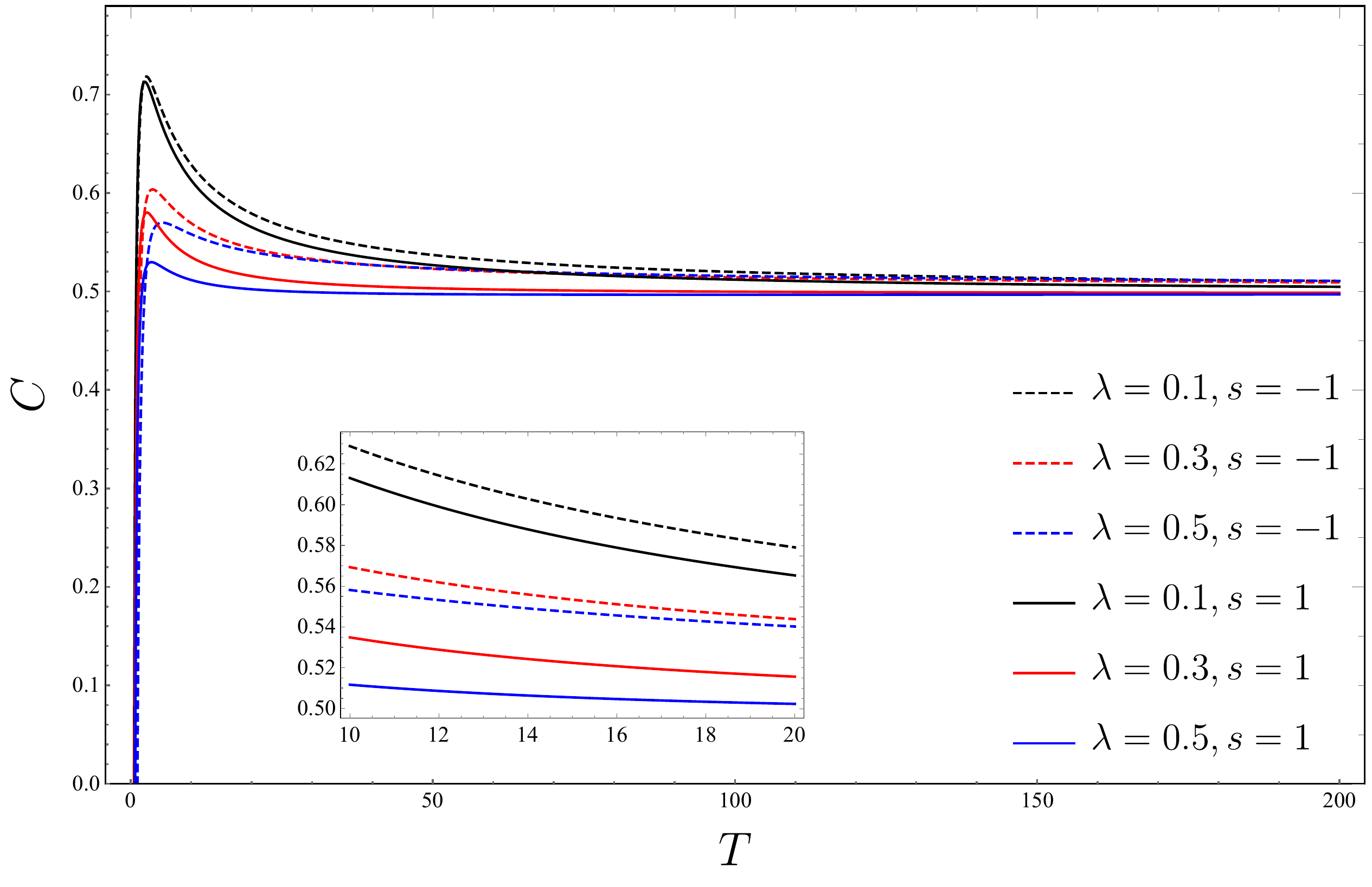}
       \subcaption{ specific heat}
       \label{fig:r333}
    \end{subfigure}
\caption{ The effect of the energy-dependence parameter $\lambda$ on thermodynamic properties. Here, we use $\gamma=0.1$, $\gamma=0.1$.}
\label{fig:rho111}
\end{figure}

To investigate the thermodynamic behavior of the energy-dependent harmonic
oscillator within the generalized Dunkl framework, we introduce the
canonical partition function,%
\begin{equation}
Z=\sum_{n=0}^{\infty }e^{-\beta \left( E_{n}^{s}-E_{0}^{s}\right) }=e^{\frac{%
2\beta \mu \alpha }{\lambda }\left[ \mu \alpha +\sqrt{1+\mu ^{2}\alpha ^{2}}%
\right] }\sum_{n=0}^{\infty }e^{-\frac{2\beta \mu \left( n+\alpha \right) }{%
\lambda }\left[ \mu \left( n+\alpha \right) +\sqrt{1+\mu ^{2}\left( n+\alpha
\right) ^{2}}\right] },
\end{equation}

where $\beta =\frac{1}{KT}$, $K$ denoting the Boltzmann constant, and 
\begin{equation}
\alpha =\frac{\eta +1-\frac{s}{2}}{2},\mu =\lambda \hbar \omega \sqrt{\left(
1-\gamma ^{2}\right) }.
\end{equation}%
Since the infinite series cannot be evaluated in closed form, the
Euler--Maclaurin summation formula is employed to approximate the summation,

\begin{equation}
\sum_{n=0}F\left( n\right) =\frac{1}{2}F\left( 0\right) +\int_{0}^{+\infty
}F\left( x\right) dx-\sum_{p=1}\frac{B_{2p}}{\left( 2p\right) !}F^{\left(
2p-1\right) }\left( 0\right) ,  \label{integral}
\end{equation}%
where $B_{2p}$ are the Bernoulli numbers and $F^{\left( 2p-1\right) }\left(
0\right) $ denotes the ($2p-1$)-th derivative of $F\left( x\right) $ at $x=0$%
. To evaluate the integral term in (\ref{integral}), we perform the
successive variable transformations%
\begin{equation}
y=\left( x+\alpha \right) \text{ and }t=\mu y+\sqrt{1+\mu ^{2}y^{2}},
\end{equation}

which convert the integral into%
\begin{equation*}
\int_{0}^{+\infty }F\left( x\right) dx=\frac{e^{\frac{\beta }{\lambda }%
+\beta E_{0}^{s}}}{2\mu }\int_{\mu \alpha +\sqrt{1+\mu ^{2}\alpha ^{2}}%
}^{+\infty }\exp \left[ -\frac{\beta }{\lambda }t^{2}\right] \left( 1+\frac{1%
}{t^{2}}\right) dt.
\end{equation*}%
After carrying out the integration, one obtains

\begin{equation}
\int_{0}^{+\infty }F\left( x\right) dx=\frac{e^{\frac{\beta }{\lambda }%
+\beta E_{0}^{s}}}{2\mu }\left[ \sqrt{\frac{\pi \beta }{\lambda }}\left( 1-%
\frac{\lambda }{2\beta }\right) \left[ \text{erf}\left( \sqrt{\frac{\beta }{%
\lambda }}\left( \alpha \mu +\sqrt{\alpha ^{2}\mu ^{2}+1}\right) \right) -1%
\right] +\frac{e^{-\frac{\beta \left( \alpha \mu +\sqrt{\alpha ^{2}\mu ^{2}+1%
}\right) ^{2}}{\lambda }}}{\alpha \mu +\sqrt{\alpha ^{2}\mu ^{2}+1}}\right] .
\label{integral28}
\end{equation}%
Retaining the Euler--Maclaurin expansion up to $p=2$, and combining Eqs. (%
\ref{integral}) and (\ref{integral28}), the partition function can be
expressed as
\begin{eqnarray}
Z &=&\frac{1}{2}+\frac{e^{\frac{2\beta \mu \alpha \mathcal{A}}{\lambda }+%
\frac{\beta }{\lambda }}}{2\mu }\left[ \left( 1-\frac{\lambda }{2\beta }%
\right) \sqrt{\frac{\pi \beta }{\lambda }}\left[ \text{erf}\left( \mathcal{A}%
\sqrt{\frac{\beta }{\lambda }}\right) -1\right] +\frac{e^{-\frac{\beta 
\mathcal{A}^{2}}{\lambda }}}{\mathcal{A}}\right]   \notag \\
&&+\frac{\beta \mu }{6\lambda }\left[ \frac{1+2\alpha ^{2}\mu ^{2}}{\mathcal{%
B}}+2\alpha \mu \right] +\frac{1}{720}\Bigg[-\left( \frac{2\beta \mu }{%
\lambda }\right) ^{3}\left( \frac{1+2\alpha ^{2}\mu ^{2}}{\mathcal{B}}%
+2\alpha \mu \right) ^{3}+  \notag \\
&&+3\left( \frac{2\beta \mu }{\lambda }\right) ^{2}\left( 2\mu +\frac{%
3\alpha \mu ^{2}}{\mathcal{B}}-\frac{\alpha ^{3}\mu ^{4}}{\mathcal{B}^{3}}%
\right) \left( \frac{1+2\alpha ^{2}\mu ^{2}}{\mathcal{B}}+2\alpha \mu
\right)   \notag \\
&&-\frac{2\beta \mu }{\lambda }\left( \frac{3\mu ^{2}}{\mathcal{B}}-\frac{%
6\alpha ^{2}\mu ^{4}}{\mathcal{B}^{3}}+\frac{3\alpha ^{4}\mu ^{6}}{\mathcal{B%
}^{5}}\right) \Bigg],
\end{eqnarray}

where 
\begin{equation}
\mathcal{A}=\left( \sqrt{\alpha ^{2}\mu ^{2}+1}%
+\alpha \mu \right) ,\mathcal{B}=\sqrt{\alpha ^{2}\mu ^{2}+1}.
\end{equation}
The partition function allows us to determine the thermodynamic properties
of the energy-dependent harmonic oscillator within the generalized Dunkl
framework. In particular, the internal energy $U$, the entropy $S$, and the
specific heat $C$ are obtained from the following relations:%
\begin{equation}
U=-\frac{\partial \ln Z}{\partial \beta },\frac{S}{K}=\ln Z-\beta \frac{%
\partial \ln Z}{\partial \beta }\text{ and }\frac{C}{K}=\beta ^{2}\frac{%
\partial ^{2}\ln Z}{\partial \beta ^{2}}.
\end{equation}
We now discuss the numerical results. Firstly, we should mention that, in all the figures, we have used $k=\omega=\hbar=m=1$. Figures \ref{fig:rho1}--\ref{fig:rho111} display the temperature dependence of the entropy, internal energy, and specific heat for different values of the Wigner parameter $\nu$, the deformation parameter $\gamma$, and the energy-dependence parameter $\lambda$. Increasing $\nu$ decreases the entropy while increasing the internal energy and specific heat, reflecting the upward shift of the energy spectrum. In contrast, increasing $\gamma$ enhances all three thermodynamic quantities, indicating that the deformation parameter strongly affects the thermal response of the system. On the other hand, increasing $\lambda$ reduces the entropy, internal energy, and specific heat owing to the energy-dependent interaction. Moreover, the entropy is always larger for the even-parity sector, whereas the internal energy and the specific heat remain higher for the odd-parity sector, in agreement with the parity splitting of the energy spectrum. \\
These results show that the generalized Dunkl deformation and the energy-dependent interaction significantly influence both the spectral and thermodynamic properties of the system, providing additional control over its thermal behavior.
\section{Conclusion}

\label{sec5}
In this work, we have developed a generalized Dunkl quantum framework incorporating energy-dependent interactions through a two-parameter deformation of the Dunkl differential operator. This extension modifies the underlying algebraic structure and naturally leads to a generalized continuity equation together with a modified probability density, conserved current, and normalization condition, ensuring a consistent probabilistic interpretation for energy-dependent quantum systems.

As an exactly solvable application, we investigated the harmonic oscillator with a linearly energy-dependent potential. Exact analytical solutions for both the eigenfunctions and energy spectrum were obtained in closed form. The spectrum exhibits several distinctive features arising from the interplay between the generalized Dunkl operator and the energy-dependent interaction, including parity splitting between even and odd states, nonlinear dependence on the quantum number, and increasing level spacing at higher energies. The limiting cases correctly recover the generalized Dunkl oscillator, the conventional energy-dependent oscillator, and the standard one-dimensional harmonic oscillator, confirming the consistency of the formalism.

The thermodynamic behavior of the system was analyzed by constructing the canonical partition function within the Euler–Maclaurin approximation. The corresponding entropy, internal energy, and specific heat reveal that the Wigner parameter, deformation parameter, and energy-dependence parameter significantly modify the thermal response of the system. In particular, the Wigner parameter increases the energy spectrum and enhances the internal energy and specific heat, the deformation parameter strongly affects all thermodynamic quantities, and the energy-dependent interaction provides an additional mechanism for controlling both spectral and thermal properties.

The present model establishes a unified framework combining generalized Dunkl operators with energy-dependent quantum interactions while preserving exact solvability. The results enrich the family of solvable quantum systems possessing reflection symmetry and may provide a useful foundation for future investigations of more general potentials, higher-dimensional Dunkl systems, relativistic extensions, supersymmetric formulations, and quantum statistical applications.

\section*{Declarations}
\subsection*{Conflicts of Interest/Competing Interests} 
The authors declare that they have no known competing financial interests or personal relationships that could have appeared to influence the work reported in this paper.

\subsection*{Data Availability Statement}
Data sharing is not applicable to this article as no new data were created or analyzed in this study

This work is supported by the Ministry of Higher Education and Scientific
Research, Algeria under the code: B00L02UN040120230003.

\end{document}